\documentclass[twocolumn]{aastex63}
\usepackage[utf8]{inputenc}

\usepackage{natbib}
\usepackage{graphicx}
\usepackage{amsmath}
\usepackage{booktabs}
\usepackage{longtable}
\usepackage{xspace}
\usepackage{hyperref}
\usepackage[T1]{fontenc}
\usepackage{lipsum}
\usepackage{comment}
\usepackage{xcolor}
\usepackage{multirow}
\usepackage{enumitem}

\usepackage{IEEEtrantools}

\usepackage{makecell}
\setcellgapes[t]{3pt}
\setcellgapes[b]{1pt}

\usepackage{soul} 

\newcommand{\msun}{\ensuremath{M_{\odot}}}

\providecommand{\mj}{\ensuremath{\,M_{\rm J}}}

\newcommand{\rearth}{\mbox{$R_{\ensuremath{\oplus}}$}}
\newcommand{\mearth}{\mbox{$M_{\ensuremath{\oplus}}$}}

\newcommand{\neff}{n_{\star, \text{eff}}}

\newcommand{\RosenthalPdgXFe}{$5.0^{+1.6}_{-1.3}\%$}
\newcommand{\RosenthalPdgFe}{$14.3^{+2.0}_{-1.8}\%$}

\newcommand{\BryanPdgcsXFe}{$4.5^{+2.6}_{-1.9}\%$}
\newcommand{\BryanPdgcsFe}{$28.0^{+4.9}_{-4.6}\%$}

\newcommand{\DefaultPdgFe}{$16.4^{+2.5}_{-2.4} \%$}
\newcommand{\DefaultPdgcsFe}{$14^{+12}_{-8} \%$}
\newcommand{\SmalldgPdgFe}{$20.2^{+2.8}_{-2.7} \%$}
\newcommand{\SmalldgPdgcsFe}{$31^{+14}_{-12} \%$}
\newcommand{\BigcsSmalldgPdgFe}{$20.2^{+2.9}_{-2.6} \%$}
\newcommand{\BigcsSmalldgPdgcsFe}{$34^{+13}_{-12} \%$}
\newcommand{\SmallcsSmalldgPdgFe}{$20.2^{+2.9}_{-2.7} \%$}
\newcommand{\SmallcsSmalldgPdgcsFe}{$47^{+17}_{-17} \%$}

\begin{document}

\title{No evidence for a metallicity-dependent enhancement of distant giant companions to close-in small planets in the California Legacy Survey}

\author[0000-0002-4290-6826]{Judah Van Zandt}
\affiliation{Department of Physics \& Astronomy, University of California Los Angeles, Los Angeles, CA 90095, USA}

\author[0000-0003-0967-2893]{Erik A Petigura}
\affiliation{Department of Physics \& Astronomy, University of California Los Angeles, Los Angeles, CA 90095, USA}

\begin{abstract}
Understanding the relationship between close-in small planets (CS) and distant giants (DG) is central to understanding the formation of planetary systems like our own. Most studies of this connection have found evidence for a positive correlation, though significant statistical and systematic uncertainties remain due to differences in sample size, target selection bias, and even the definitions of `close-in small' and `distant giant' planets. Recently, \cite{BryanLee2024} conducted a study of 184 stars hosting super-Earths ($M \sin i = 1-20 \, \mearth$ or $R = 1-4 \, \rearth$) to determine the effect of stellar metallicity on the prevalence of distant giant companions ($M \sin i = 0.5-20 \, \mj$, $a = 1-10$ AU). They found that such giants are twice as common in the presence of inner planets, but \textit{only} in metal-rich systems: P(DG|CS, [Fe/H]>0) = \BryanPdgcsFe \xspace vs. P(DG|[Fe/H]>0) = \RosenthalPdgFe. Further, they found that this correlation disappears for metal-poor stars: P(DG|CS, [Fe/H]$\leq$0) = \BryanPdgcsXFe \xspace vs. P(DG|[Fe/H]$\leq$0) = \RosenthalPdgXFe. We conducted an analogous study to test whether this effect was present in the California Legacy Survey sample. Using the same planet definitions, we did not find evidence for an enhancement in metal-rich systems: P(DG|CS, [Fe/H]>0) = \DefaultPdgcsFe \xspace vs. P(DG|[Fe/H]>0) = \DefaultPdgFe. We also found a $2\sigma$ tension between our conditional rate and a 2x enhancement. We brought our results into closer agreement with \cite{BryanLee2024} by repeating our analysis with different planet definitions. For example, considering only CS planets with $1 \leq M\sin i \leq 10 \, \mearth$ \xspace and DG planets with  $0.5 \leq M\sin i \leq 20 \, \mj$ gives P(DG|CS, [Fe/H]>0) = \SmallcsSmalldgPdgcsFe~ and P(DG|[Fe/H]>0) = \SmallcsSmalldgPdgFe, consistent with a 2x enhancement. However, we find that this 2x enhancement is also present when we apply the same planet definitions to the overall sample, challenging the idea that metallicity is responsible. Our sample, though only $\sim1/6$ the size of \cite{BryanLee2024}'s, was assembled and analyzed self-consistently. The discrepancy between our findings and theirs may arise from a variety of sources, underscoring the need for large exoplanet surveys to overcome both small number statistics and sample inhomogeneities.

\end{abstract}

\section{Introduction}
\label{sec:introduction}
Planets between the size of Earth and Neptune with orbits less than 1 AU are ubiquitous \citep{Petigura2018}. In contrast, radial velocity (RV) surveys have uncovered a separate population of giant planets at wider separations ($a \approx 1-10$ AU), which orbit $10-20\%$ of stars, depending on the definition used (\citealt{Rosenthal2021, Wittenmyer2020, Fischer2014, Cumming2008}). Although rarer than the super-Earths and sub-Neptunes, these giants are believed to have a significant impact on the development of close-in planets (\citealt{BatyginLaughlin2015, Izidoro2015}).

The relationship between close-in small planets (CS) and outer distant giants (DG) has garnered significant attention in recent years. \cite{ZhuWu2018} and \cite{Bryan2019} assembled samples of stars with archival RVs to calculate the occurrence of distant giants in systems hosting close-in small planets. Both studies found that distant giant occurrence is enhanced in the presence of inner small planets (P(DG|CS) = $32 \pm 8\%$ and $39 \pm 7\%$, respectively). \cite{Rosenthal2022} leveraged the California Legacy Survey (CLS), a large sample of FGKM RV targets with multi-decade baselines, to answer the same question, and also found an enhancement: P(DG|CS) = $41^{+15}_{-13}\%$. 

The emerging picture of a positive correlation was challenged by \cite{Bonomo2023}, who found evidence for an anti-correlation (P(DG|CS) = $9.3^{+7.7}_{-2.9}\%$) in a sample of 38 $Kepler$ and $K2$ systems. RV surveys tailored to the detection of outer giants are currently being carried out on samples with transiting inner planets discovered by $Kepler$ \citep{Weiss2024} and $TESS$ \citep{VanZandt2023}.

\cite{Zhu2024} added another dimension to the DG-CS relation by proposing that the enhancement of giants found by earlier studies is driven by host star metallicity. In particular, they suggested that giants are strongly enhanced in the presence of inner companions \textit{only} in systems with metal-rich ([Fe/H]>0) host stars. They demonstrated that analyzing only the metal-rich stars in the \cite{Bonomo2023} sample produces a positive correlation.

To test the claim put forward by \cite{Zhu2024}, \cite{BryanLee2024} compiled a sample of 184 Sun-like stars with archival RV observations, as well as transit- and RV-discovered CS planets. They found that distant giants are twice as prevalent in the presence of inner small planets versus the field rate for metal-rich systems, reinforcing the hypothesis that metallicity plays a key role in this connection.

The enhancement in giant planet occurrence around metal-rich stars is well known (e.g., \citealt{Santos2004, Fischer2005, Johnson2010}). Thus, one would expect that in metal-rich systems, the prevalence of DG companions in CS systems would increase, but so would the prevalence of DG planets overall, nullifying any relative increase. The conclusions of \cite{Zhu2024} and \cite{BryanLee2024} imply that the conditional occurrence rate of DGs is a more steeply rising function of metallicity than the field occurrence rate. 

In this paper we emulate the study carried out by \cite{BryanLee2024} using the California Legacy Survey (CLS) sample, a catalog of 719 FGKM stars and 178 planets discovered through RV monitoring over 30 years. Our objective is to test the claim that there exists a strong positive correlation between CS and DG planets specifically for metal-rich host stars in this sample. We review the methods and results of \cite{BryanLee2024} in Section \ref{sec:bryan_lee}. In Section \ref{sec:cls} we give an overview of the CLS sample. We describe our binomial occurrence model in Section \ref{sec:occurrence_model}, and use it to calculate giant planet occurrence rates in Section \ref{sec:occurrence_calculation}. We explore the effects of expanding our planet definitions in Section \ref{sec:broad_defs}, and discuss our results in Section \ref{sec:discussion}.

\section{Bryan \& Lee (2024)}
\label{sec:bryan_lee}

This paper examines the key finding of \cite{BryanLee2024}, hereafter BL24, namely, that in metal-rich systems, DG planets are twice as prevalent in the presence of CS planets as the field DG rate for metal-rich systems in the CLS sample. Therefore, we begin by describing the BL24 sample, including its construction and planet class definitions.

BL24 selected 184 FGK stars with >20 publicly available RVs and hosting at least one CS planet, which they defined as having either $M \sin i = 1-20 \, \mearth$ or $R = 1-4 \, \rearth$. Figure \ref{fig:sample_comparison} shows the minimum masses and separations of the planets in their sample.\footnote{We cross-referenced the system names provided in Table A.1 of BL24 with the Exoplanet Archive (https://exoplanetarchive.ipac.caltech.edu/; Accessed 9 June 2024) to reconstruct their sample. We found only one meaningful discrepancy: the putative super-Earth HD 26965 b was initially reported by \cite{Ma2018}, but recently found to be spurious \citep{Laliotis2023}. The removal of this planet would also exclude HD 26965 from the BL24 sample. We anticipate that this change would have a negligible effect on the statistical findings of BL24, and conduct our comparison against their original results.}


They calculated the conditional occurrence rate of DG planets, which they defined by $M \sin i = 0.5-20 \, \mj$ and $a = 1-10$ AU, in the metal-rich and metal-poor subsamples, finding that P(DG|SE, [Fe/H]>0) = \BryanPdgcsFe \xspace and P(DG|SE, [Fe/H]<0) = \BryanPdgcsXFe.

They then calculated the distant giant field occurrence rates using the CLS sample of \cite{Rosenthal2021}, hereafter R21. They first removed 121 M dwarfs ($M_{\star}< \, 0.6 \, \msun$) and four targets with fewer than 20 RV observations, and found P(DG|[Fe/H]>0) = \RosenthalPdgFe \xspace and P(DG|[Fe/H]<0) = \RosenthalPdgXFe~ among the remaining 594 stars. We refer to this filtered subset as the R21 sample in the remainder of this paper, and we show the planets orbiting these stars in Figure \ref{fig:sample_comparison}. Whereas the conditional and field occurrence rates for metal-poor systems are comparable, the metal-rich conditional rate is larger than its field counterpart by a factor of two. The sharp divide between the metal-rich and metal-poor samples supports the claim of \cite{Zhu2024}.

\section{The California Legacy Survey}
\label{sec:cls}
In this work, we derive all conditional and field occurrence rates using the same 594-star subset of the California Legacy Survey sample used by BL24. Using a single stellar and planetary sample naturally controls for selection biases when comparing different statistics. This sample consists of FGK stars with RV baselines spanning up to 30 years. The R21 sample has a median metallicity of 0.0 dex and a median stellar mass of 0.94 $M_{\odot}$. The R21 catalog comprises 171 substellar and planetary companions ($M \sin i$ < 80 $\mj$), 154 of which have $M \sin i$ < 13 $\mj$, and 33 of which have $M \sin i$ < 30 $\mearth$. All companions in the sample were detected using a uniform search algorithm (RVSearch, \citealt{Rosenthal2021}). \cite{Fulton2021} measured the sensitivity of this algorithm in the R21 data set through a suite of injection/recovery experiments published as part of that study.

\begin{figure*}
  \centering\includegraphics[width=1\textwidth]{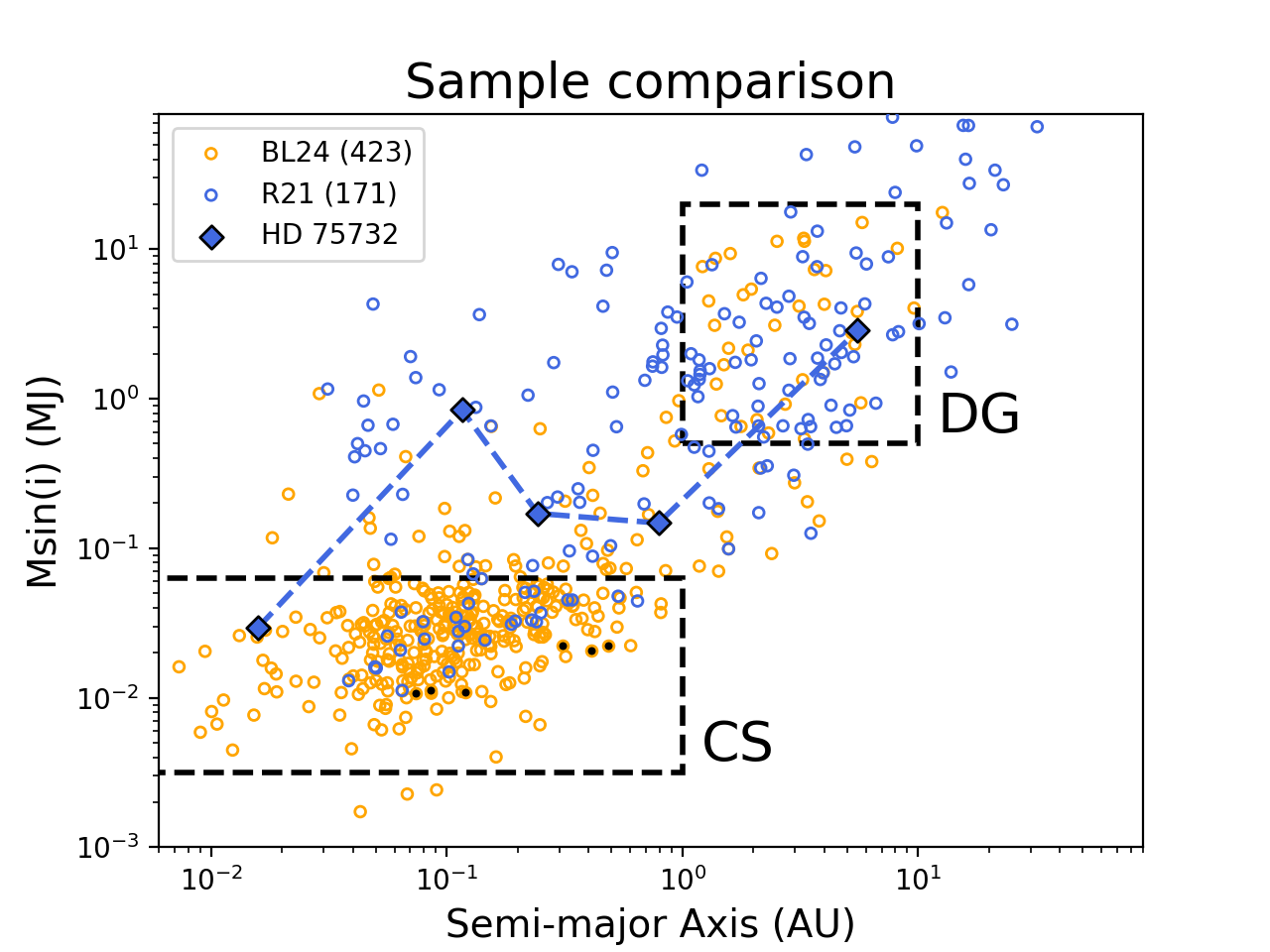}
 \centering\caption{The companions in the sample \cite{BryanLee2024} (yellow) and the FGK subsample of \cite{Rosenthal2021} (blue). BL24 has companions between $\sim1$ \mearth \xspace and 20 \mj, and R21 includes companions from a few \mearth \xspace up to 80 \mj. The legend gives the number of companions in each sample in parentheses. We show BL24's definitions of close-in small (CS) and distant giant (DG) planets with black dashed boxes. We use blue diamonds to highlight HD 75732, the only system in the metal-rich R21 sample with planets satisfying both of these definitions. We use filled black circles to show six planets in the BL24 sample without measured masses. We estimate their masses using the mass-radius relationship of \cite{Weiss2014}. The distributions of giant plants are similar among the two samples. They also have comparable inner planet distributions, though BL24 contains many more such planets due to its selection function. This latter similarity makes it unlikely that a dependence of giant occurrence on inner planet properties, if such a dependence exists, affected our comparison.}
  \label{fig:sample_comparison}
\end{figure*}

The CLS sample may not be free of bias, as \cite{Zhu2022} showed that it hosts hot Jupiters at about three times the canonical occurrence rate of $1\%$ \citep{Wright2012}. Nevertheless, our goal is not to measure the absolute conditional occurrence of DG planets. Rather, it is to test BL24's finding that  metal-rich stars exhibit an increased relative enhancement of P(DG|CS) over P(DG). To do this, we calculated P(DG) and P(DG|CS) from a single sample, allowing us to rule out sample differences as a source of disparity. We chose the R21 sample to match BL24 as closely as possible. We show in Figure \ref{fig:sample_comparison} that the distributions of small inner planets and outer giants are comparable in the BL24 and R21 samples.

\section{Binomial Occurrence Model}
\label{sec:occurrence_model}

In this work, we define occurrence as the fraction of systems hosting at least one planet, rather than the number of planets per star. Following BL24, we calculated occurrence using binomial statistics, assuming that planet occurrence was $\log$-uniform over the parameter space we considered ($M \sin i = 0.5-20 \, \mj$, $a = 1-10$ AU). We calculated the average completeness $Q$ of the R21 survey and used that to calculate $\neff = Q \cdot n_{\star}$, where $n_{\star}$ is the total number of stars in the sample and $\neff$ is the effective number of stars after correcting for completeness. The binomial likelihood function for the fraction $f_{\text{cell}}$ of stars hosting planets in a given domain is

\begin{equation}
    P(f_{\text{cell}} |n_{\text{det}}, \neff) =  C f_{\text{cell}}^{n_{\text{det}}}(1-f_{\text{cell}})^{n_{\text{nd}}},
\end{equation}

\noindent where $n_{\text{det}}$ is the number of systems hosting a detected planet, $n_{\text{nd}} = \neff - n_{\text{det}}$ is the number of systems in which no planet was detected, and $C$ is a normalization constant:
\footnote{We use $\Gamma$ to represent the extension of the factorial function to real numbers: $\Gamma(x+1) = x!$.}

\begin{equation}
    C =  \frac{(\neff+1) \cdot \Gamma(\neff+1)}{\Gamma(n_{\text{det}}+1) \Gamma(n_{\text{nd}}+1)}.
\end{equation}

\noindent We set a uniform prior between zero and one on the planet occurrence rate and numerically optimized our likelihood function.

\section{No Evidence of Enhanced Occurrence of Distant Giants in Metal-rich Stars}
\label{sec:occurrence_calculation}
\subsection{P(DG)}
We first calculated the field occurrence of distant giants in the R21 sample to establish a point of comparison for conditional occurrence. Of the 594 stars in the R21 sample, 52 host one or more distant giants, and 43 of those hosts are metal-rich. We calculated metal-rich and metal-poor distant giant field occurrence rates of P(DG|[Fe/H]>0) = \DefaultPdgFe~ and P(DG|[Fe/H]$\leq$0) = $5.8^{+2.0}_{-1.7} \%$. 

Our metal-rich field rate differs from that obtained by BL24 (\RosenthalPdgFe) by $>1\sigma$. For two unrelated studies of the same phenomenon, such a disparity could reasonably be attributed to statistical variation. However, we used the same sample, planet definitions, and occurrence model as BL24. We therefore attribute this disagreement to our different completeness correction methodologies. In brief, both our and BL24's completeness corrections involved injecting and recovering planetary signals in the measured RVs. However, BL24 considered a signal ``recovered'' if a one-planet model offered a significantly improved fit over the flat line model ($\Delta$BIC$>10$). By contrast, R21 required that a planetary model be preferred and that the recovered parameters agree with the injected ones to within 25\%. This stricter criterion would lead to lower completeness, and therefore a higher measured occurrence, as we observe.

In principle, there is no problem with using a more or less strict completeness correction. It is only necessary that the same criteria be applied during the initial planet search and the completeness correction. Thus, a strict planet search will find fewer planets and correct more, while a less strict search will find more planets and correct less. Because the R21 sample was compiled using RVSearch, we expect that the RVSearch completeness framework we used provides a more accurate correction.

\subsection{P(DG|CS)}
\begin{figure*}
  \centering\includegraphics[width=1\textwidth]{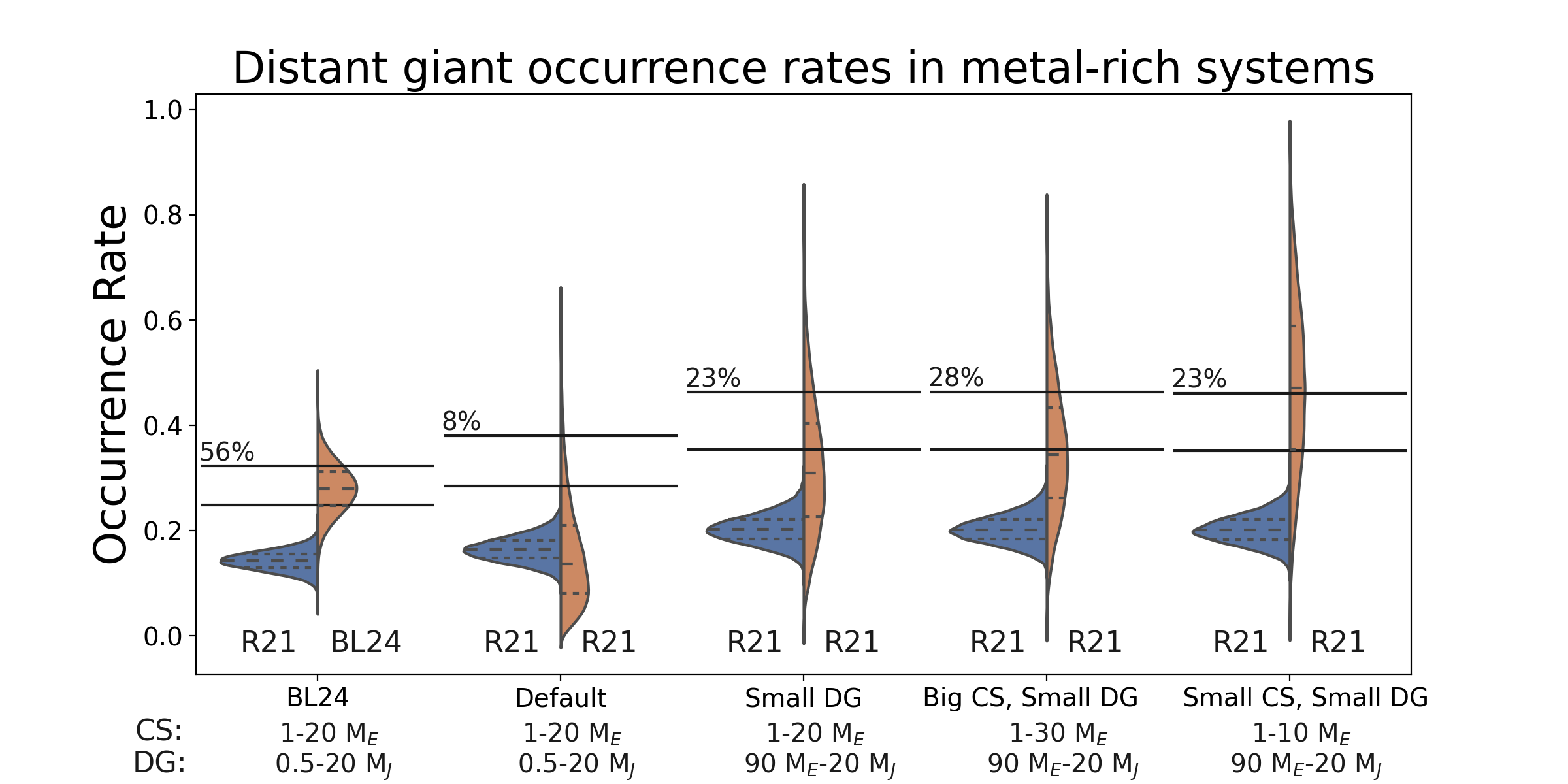}
 \centering\caption{Measurements of field (blue) and conditional (orange) occurrence of distant giants in metal-rich subsets of the \cite{BryanLee2024} and \cite{Rosenthal2021} samples. BL24 found that distant giants are about twice as prevalent in the presence of close-in small planets as the field rate they measured from the R21 sample (far left). We measured both the conditional and field rates using the R21 sample and four different sets of definitions for close-in small planets and distant giants. The "Default" definition matches BL24: the inner planets satisfy $1 \leq M\sin i \leq 20 \, \mearth$ \xspace and the distant giants satisfy  $0.5 \leq M\sin i \leq 20 \, \mj$ and $1 \leq a \leq 10$ AU. The "Small DG" definition used a $90 \, \mearth$ floor for distant giants. The "Big CS, Small DG" definition used a $30 \, \mearth$ ceiling for inner planets and a $90 \, \mearth$ floor for distant giants. Finally, our "Small CS, Small DG" definition used a $10 \, \mearth$ ceiling for inner planets and a $90 \, \mearth$ floor for distant giants. We annotate each distribution with the planet sample used to calculate it, and include the CS and DG definitions on the $x$-axis. Horizontal dotted lines denote distribution quartiles. We use black solid lines to indicate the 95\% confidence interval of a 2x enhancement over each field occurrence rate, and label the upper line with the percentage of each conditional occurrence distribution that falls within it. BL24's conditional occurrence distribution is fully consistent with a 2x enhancement, whereas only 8\% of our Default distribution overlaps this interval. Expanding our definitions of inner planets and distant giants increases the conditional rate, because a significant fraction of the giant planets in the R21 sample have masses below $0.5 \mj$.}
  \label{fig:pdgcs_plot}
\end{figure*}

We next calculated the conditional occurrence of distant giants in the presence of close-in small planets (CS). We adopted the mass limits used by BL24 for these planets, $M \sin i = 1-20 \, \mearth$. We also used a separation limit of $a<1$ AU, though this filter did not remove any targets from our subsample. Under this definition, 18 R21 systems host a close-in small planet, and only one hosts both a close-in small planet and a distant giant. Of this sub-sample, 13 systems, including the distant giant host, are metal-rich. We calculated a metal-rich distant giant conditional occurrence rate of P(DG|CS, [Fe/H]>0) = \DefaultPdgcsFe. This value is \textit{lower} than our metal-rich field occurrence rate calculated above, in tension with most previous studies of this relationship using various CS and DG definitions (e.g. \citealt{Bryan2019, Rosenthal2022, Zhu2024}). Moreover, our measured rate is inconsistent with a 2x enhancement at  the 2$\sigma$ level, in contrast to BL24's finding of a 2x enhancement at 3$\sigma$ confidence. We compare these measurements and their uncertainties in Figure \ref{fig:pdgcs_plot}.

\begin{deluxetable*}{llccc}
\tablewidth{0pt}
\caption{\textsc{Occurrence Rates}}
\label{tab:occurrence_rates}

\tablehead{
\colhead{} & \colhead{Metallicity} & \colhead{} & \colhead{} & \colhead{\% within 2$\sigma$} \vspace{-0.2cm}\\
\colhead{Test} & \colhead{Criterion} & \colhead{P(DG|CS)} & \colhead{P(DG)} & \colhead{(2x Enhancement)}}
\startdata
           BL24    &        [Fe/H]>0 &  $28.0^{+4.9}_{-4.6}\%$ & $14.3^{+2.0}_{-1.8}\%$ &  --- \\
                   & [Fe/H] $\leq$ 0 &   $4.5^{+2.6}_{-1.9}\%$ &  $3.8^{+1.4}_{-1.1}\%$ & --- \\
           Default &        [Fe/H]>0 &  $14^{+12}_{-8}\%$ & $16.4^{+2.5}_{-2.4}\%$ &  8\% \\
                   & [Fe/H] $\leq$ 0 &                  <45\% &  $5.8^{+2.0}_{-1.7}\%$ & 22\% \\
                   &             --- &   $10^{+9}_{-6}\%$ & $12.0^{+1.7}_{-1.6}\%$ &  9\% \\
          Small DG &        [Fe/H]>0 & $31^{+14}_{-12}\%$ & $20.2^{+2.8}_{-2.7}\%$ & 23\% \\
                   & [Fe/H] $\leq$ 0 &                  <46\% &  $6.3^{+2.3}_{-1.8}\%$ & 23\% \\
                   &             --- &  $23^{+11}_{-9}\%$ & $14.6^{+2.0}_{-1.9}\%$ & 22\% \\
  Big CS, Small DG &        [Fe/H]>0 & $34^{+13}_{-12}\%$ & $20.2^{+2.9}_{-2.6}\%$ & 28\% \\
                   & [Fe/H] $\leq$ 0 &                  <38\% &  $6.3^{+2.2}_{-1.9}\%$ & 24\% \\
                   &             --- &  $25^{+11}_{-9}\%$ & $14.5^{+1.9}_{-1.8}\%$ & 25\% \\
Small CS, Small DG &        [Fe/H]>0 & $47^{+17}_{-17}\%$ & $20.2^{+2.9}_{-2.7}\%$ & 23\% \\
                   & [Fe/H] $\leq$ 0 &                  <72\% &  $6.2^{+2.2}_{-1.8}\%$ & 16\% \\
                   &             --- & $40^{+16}_{-15}\%$ & $14.5^{+2.0}_{-1.8}\%$ & 16\% \\
\enddata
\tablecomments{(1) BL24: CS = 1--20 \mearth; DG = 1--10 AU, 0.5--20 \mj; (2) Default: Same as BL24 definition; (3) Small DG: Same as BL24 but with DG mass limits of $90 \, \mearth$--$20 \, \mj$; (4) Big CS, Small DG: CS = 1--30 \mearth; DG = $90 \, \mearth$--$20 \, \mj$; (5) Small CS, Small DG: CS = 1--10 \mearth; DG = $90 \, \mearth$--$20 \, \mj$. We provide the median occurrence rates, with uncertainties representing 16th and 84th percentiles. For occurrence rates consistent with zero, we quote 90\% upper limits. The right-most column indicates the fraction of each of our conditional posterior distributions that falls within the 95\% confidence interval of a 2x-enhancement over the corresponding field distribution.}
\end{deluxetable*}

\section{Broader Planet Definitions Increase P(DG|CS)}
\label{sec:broad_defs}
We performed our analysis in the previous sections in exact analogy to BL24, including using the same definitions of close-in small planets and distant giants. Under these definitions, there were 43 DG hosts, 13 CS hosts, and 1 system hosting both. Here we examine the effect of broadening these criteria to admit a wider range of planets. The values we quote in this section are specifically for the metal-rich sample unless otherwise indicated. We show the metal-rich occurrence rate distributions in Figure \ref{fig:pdgcs_plot}, and we list the corresponding results for the full and metal-poor samples in Tables \ref{tab:occurrence_rates} and \ref{tab:sample_size_completeness}.

First, we decreased the lower bound on DG mass to 90 $\mearth$ to include Saturn-mass planets, giving 48 DG hosts, 13 CS hosts, and 3 hosts of both. The decreased mass limit added more DGs to our sample and also extended the parameter space to regions of lower sensitivity, resulting in a larger completeness correction. As a result, we found enhancements in both the field rate and the conditional rate, P(DG|[Fe/H]) = \SmalldgPdgFe~ and P(DG|CS, [Fe/H]) = \SmalldgPdgcsFe, this time finding consistency with a 2x enhancement within 1$\sigma$.

Next, we retained the lower DG mass bound increased the upper bound on CS planets to 30 $\mearth$. We recovered the same 48 DG hosts, this time with 16 CS hosts among them, and 5 systems hosting both. Our results under this change, P(DG|[Fe/H]) = \BigcsSmalldgPdgFe~ and P(DG|CS, [Fe/H]) = \BigcsSmalldgPdgcsFe, were consistent with a 2x enhancement. However, the overall sample exhibited a comparable relative enhancement under these same definitions, suggesting that metallicity is not responsible (see Table \ref{tab:occurrence_rates}).

Finally, we lowered the CS mass ceiling to 10 $\mearth$ to test giant occurrence around typical super-Earths. This definition gave 48 DG hosts, 8 CS hosts, and 3 hosts of both. The only difference between this definition and the first definition in this section is the smaller CS mass ceiling (10 $\mearth$ versus 20 $\mearth$). Although this change cut out five CS hosts, we retained all three of the systems hosting both a CS and a DG, meaning that no giants in our sample orbit inner planets with mass $10-20 \, \mearth$. We found P(DG|[Fe/H]) = \SmallcsSmalldgPdgFe~ and P(DG|CS, [Fe/H]) = \SmallcsSmalldgPdgcsFe. In this case, our results were fully consistent with a 2x enhancement, with our median conditional rate greater than the field rate by a factor of 2.2. Again, we saw a comparable relative enhancement (2.6x) when we applied these definitions to the overall sample.

\section{Excluding evolved stars has little effect on occurrence}
\label{sec:logg_cut}
Stars hosting close-in planets are expected to engulf these companions as they evolve off the main sequence \citep{Villaver2014}. Although the number of known exceptions to this picture is growing (e.g., \citealt{Grunblatt2024, Wittenmyer2022, Wang2019}), evolved stars as a population may host CS planets at a reduced rate, and would therefore tend to increase P(DG) in relation to P(DG|CS).

We removed evolved hosts and repeated our analysis to examine conditional occurrence among only main sequence stars. To exclude evolved hosts from our sample, we removed all stars with $\log (g)$<4, which reduced our sample from 598 to 531 stars, 324 of which were metal-rich and 207 of which were metal-poor. The 67 evolved stars we removed host 12 of the 52 giant planets in the CLS, including HD 11964 A, which hosts both a transiting super-Neptune ($M\sin i \sim 24 \, \mearth$) and an outer sub-Jupiter ($M\sin i \sim 0.6 \, \mj$).

We then recomputed the statistics of Sections \ref{sec:occurrence_calculation} and \ref{sec:broad_defs}. Our average completeness did not change by more than 2\% for any of our experiments. We found that removing evolved stars decreased our field distant giant rates, though not enough to reach statistical significance. Finally, because only a single CS system was removed in our "Big CS, Small DG" experiment, and none were removed in the other experiments, our conditional occurrence rates were essentially unaffected.

\section{Discussion}
\label{sec:discussion}
We used the R21 sample to reproduce the measurement of distant giant occurrence in metal-rich systems conducted by \cite{BryanLee2024}. 
We focused on the primary result of that study: that the conditional occurrence of distant giants is significantly enhanced over the field rate specifically for metal-rich host stars. We used BL24's finding of a factor of 2 enhancement as a benchmark to test against.

Adopting the same planet definitions and statistical methods as BL24, we did not find evidence for a strong enhancement of DGs in the presence of CS planets in metal-rich systems. Instead, our measurements are in tension with a factor of 2 enhancement at the $2\sigma$ level.

We tested the effect of changing our DG and CS definitions on our inferred occurrence rates. Broadening these definitions to admit more CS or DG planets increased the measured conditional occurrence by a greater factor than it did the field occurrence (see Table \ref{tab:occurrence_rates}), leading to larger relative enhancements for broader planet definitions. However, we saw similar increases in relative enhancement for the overall sample, suggesting that the enhancement of P(DG|CS) over P(DG) is not dependent on metallicity.

As a final test, we checked whether removing evolved stars from our sample caused an appreciable change in our calculated occurrence rates. We found that although our basic prediction that this would decrease P(DG) in relation to P(DG|CS) was borne out, the effect was not statistically significant.

The design of our study differs from BL24 in both the sizes of our respective samples and the homogeneity of our analyses. BL24 assembled a sample of 184 small-planet-hosting stars, a number of which also hosted distant giants. Their large sample size allowed them to compute P(DG|CS) with a 5\% fractional uncertainty. However, because their sample only included systems with inner planets, they could not use it to establish P(DG). They therefore compared their conditional occurrence rate to the field rate from the R21 sample with the assumption that the two samples were drawn from the same underlying distribution of stars with the same P(DG). This assumption may not be correct, and the increased measurement precision from using a larger sample may have come at the cost of decreased accuracy.

We used the R21 sample to compute P(DG) and P(DG|CS) self-consistently, so we did not contend with biases arising from different samples. However, the R21 sample contains far fewer CS planets than BL24, resulting in large fractional uncertainties on our measurements of P(DG|CS).

A third difference is the use of different completeness correction algorithms. In particular, we used stricter criteria for injected planet recoveries, resulting in lower average completeness and therefore higher inferred occurrence rates. Because these criteria resemble the actual detection pipeline used to compile the R21 sample more closely than BL24's, we expect that our completeness correction is more accurate.

Despite the rapid growth of the census of exoplanets over the last few decades, samples large enough to support homogeneous, multi-dimensional analyses, such as conditional occurrence rate estimates, are not yet available. The largest blind RV surveys, with sample sizes of $\sim$1000, were compiled over decades and are unlikely to grow by an order of magnitude in the near future. However, improvements in RV precision will provide increased sensitivity to small planets over time. This will directly enable more precise measurements of P(CS), but it will also reveal CS planets in DG-hosting systems. Therefore, we expect improved precision on P(DG|CS) even in the absence of larger samples.

Leveraging other detection methods is another promising avenue for improving occurrence rate statistics. For example, using uniform transit missions to identify close-in small planets can efficiently establish a sample of CS-hosting stars, and using RVs to follow up such a sample homogeneously (i.e., without regard to metallicity, probability of a distant giant, etc.) can provide P(DG|CS) even with modest RV precision. Furthermore, searching for distant giant signatures in astrometric data from $Gaia$ could be used to make a robust measurement of P(DG) for comparison. The risk of inter-sample biases from cross-method comparisons remains a concern, but will diminish as large surveys approach statistically identical target samples.

\section{Acknowledgments}
We thank the reviewer for helpful comments which improved this manuscript.

J.V.Z. acknowledges support from NASA FINESST Fellowship 80NSSC22K1606. J.V.Z. and E.A.P. acknowledge support from NASA XRP award 80NSSC21K0598.

\begin{deluxetable*}{llccccc}
\tablewidth{0pt}
\caption{\textsc{Samples Sizes and Completeness}}
\label{tab:sample_size_completeness}

\tablehead{
\colhead{} & \colhead{} & \colhead{Metallicity} & \colhead{Average} & \colhead{} & \colhead{} & \colhead{} \vspace{-0.2cm}\\
\colhead{Test} & \colhead{Measurement} & \colhead{Criterion} & \colhead{Completeness} & \colhead{$N_{\star}$} & \colhead{$N_{\star, 
\text{eff}}$} & \colhead{$N_{\text{hosts}}$}}
\startdata
           Default &    P(DG) &        [Fe/H]>0 & 0.70 &   351 & 244.0 &    43 \\
                   & P(DG|CS) &        [Fe/H]>0 & 0.83 &    13 &  10.7 &     1 \\
                   &    P(DG) & [Fe/H] $\leq$ 0 & 0.70 &   243 & 170.8 &     9 \\
                   & P(DG|CS) & [Fe/H] $\leq$ 0 & 0.81 &     5 &   4.1 &     0 \\
                   &    P(DG) &             --- & 0.70 &   594 & 413.8 &    52 \\
                   & P(DG|CS) &             --- & 0.82 &    18 &  14.8 &     1 \\
          Small DG &    P(DG) &        [Fe/H]>0 & 0.63 &   351 & 222.6 &    48 \\
                   & P(DG|CS) &        [Fe/H]>0 & 0.81 &    13 &  10.6 &     3 \\
                   &    P(DG) & [Fe/H] $\leq$ 0 & 0.64 &   243 & 156.5 &     9 \\
                   & P(DG|CS) & [Fe/H] $\leq$ 0 & 0.80 &     5 &   4.0 &     0 \\
                   &    P(DG) &             --- & 0.64 &   594 & 377.9 &    57 \\
                   & P(DG|CS) &             --- & 0.81 &    18 &  14.6 &     3 \\
  Big CS, Small DG &    P(DG) &        [Fe/H]>0 & 0.63 &   351 & 222.6 &    48 \\
                   & P(DG|CS) &        [Fe/H]>0 & 0.81 &    16 &  13.0 &     5 \\
                   &    P(DG) & [Fe/H] $\leq$ 0 & 0.64 &   243 & 156.5 &     9 \\
                   & P(DG|CS) & [Fe/H] $\leq$ 0 & 0.83 &     6 &   5.0 &     0 \\
                   &    P(DG) &             --- & 0.64 &   594 & 377.9 &    57 \\
                   & P(DG|CS) &             --- & 0.82 &    22 &  18.0 &     5 \\
Small CS, Small DG &    P(DG) &        [Fe/H]>0 & 0.63 &   351 & 222.6 &    48 \\
                   & P(DG|CS) &        [Fe/H]>0 & 0.81 &     8 &   6.5 &     3 \\
                   &    P(DG) & [Fe/H] $\leq$ 0 & 0.64 &   243 & 156.5 &     9 \\
                   & P(DG|CS) & [Fe/H] $\leq$ 0 & 0.70 &     2 &   1.4 &     0 \\
                   &    P(DG) &             --- & 0.64 &   594 & 377.9 &    57 \\
                   & P(DG|CS) &             --- & 0.79 &    10 &   7.9 &     3 \\
\enddata
\tablecomments{$N_{\star}$ gives the number of stellar hosts satisfying the associated metallicity criterion (and the CS criterion, for P(DG|CS) measurements). $N_{\star, \text{eff}}$ gives the product of $N_{\star}$ and the completeness to distant giants among this stellar subsample. $N_{\text{hosts}}$ denotes the number of systems hosting at least one DG planet under the given criteria.}
\end{deluxetable*}

\bibliography{bib.bib}

\end{document}